\documentclass{elsart3p}

\usepackage{graphics}
\usepackage{graphicx}
\usepackage{epsfig}

\usepackage{amssymb}

\begin{document}

\begin{frontmatter}

\def \gev {GeV/$c$}
\def \pt  {p_{T}}
\def \cf {C_{6}F_{14}}
\def \snn {\sqrt{s_{NN}}}



\title{The ALICE HMPID detector ready for collisions at the LHC}


\author[INFN,KFKI]{Levente Molnar}
\ead{Levente.Molnar@ba.infn.it}
\address[INFN]{INFN~Sezione~di~Bari,~Via~E.~Orabona~4,~70126~Bari,~Italy}
\address[KFKI]{Res.~Inst.~Particle~\&~Nucl.~Phys.,~Hungarian~Academy~of~Sciences,~Hungary}
\author{On behalf of the ALICE-HMPID group}

\begin{abstract}
ALICE has been specifically optimized to study heavy-ion collisions at
the LHC, up to a charged particle density of 8000 per unit of
rapidity in central heavy-ion collisions at $\snn$ = 5.5 TeV.
The High Momentum Particle Identification Detector (HMPID) has a
proximity focusing geometry with a liquid $\rm C_{6}F_{14}$ Cherenkov
radiator coupled to Multi-Wire Pad Chambers (MWPC) equipped with
CsI photocathodes, over a
total active area of 11 $\rm m^2$. It has been designed to identify
charged pions and kaons in the range 1 $\leq p \leq$ 3 \gev~and
protons in the range 2 $\leq p \leq$ 5 \gev.
The as-built detector and all relevant subsystems (gas,
liquid $\rm C_{6}F_{14}$, cooling and control) are described.
Installation issues and first commissioning results
are also presented.
\end{abstract}

\begin{keyword}
HMPID \sep ALICE \sep RICH \sep CsI photocathode

\PACS 25.6 \sep 34.8.a \sep 29.40.Ka
\end{keyword}

\end{frontmatter}

\section{Introduction}
\label{Sec:Intro}
The ALICE experiment at CERN is optimized to study heavy-ion collisions
at center-of-mass energy $\sqrt{s_{NN}}=$ 5.5 GeV at the LHC~\cite{ALICE,ALICEPPR1}.
The High Momentum Particle Identification Detector~\cite{ALICEPPR1,HMPIDTDR}
(HMPID) in ALICE is dedicated to the inclusive measurement of charged particles,
both in heavy-ion and proton-proton
collisions. The single arm design of the HMPID covers 5\% of the
central barrel region of ALICE at mid-rapidity. The HMPID enhances
the particle identification capabilities of ALICE, extending the momentum range further
than is accessible to individual central tracking detectors: ITS, TPC and TOF.
The HMPID can identify charged pions and kaons in the momentum
range $p \sim 1-3~{\rm GeV}/c$ and protons up to $p~\sim$~5~GeV/$c$ with
3$\sigma$ separation.

This paper focuses on the as-built description of
the HMPID and the first commissioning results. Further details of HMPID hardware and
physics goals can be found in~\cite{ALICEPPR1,HMPIDTDR,ALICEPPR2}.
\begin{figure}[htbp]
    \centering
        \includegraphics[width=0.49\textwidth]{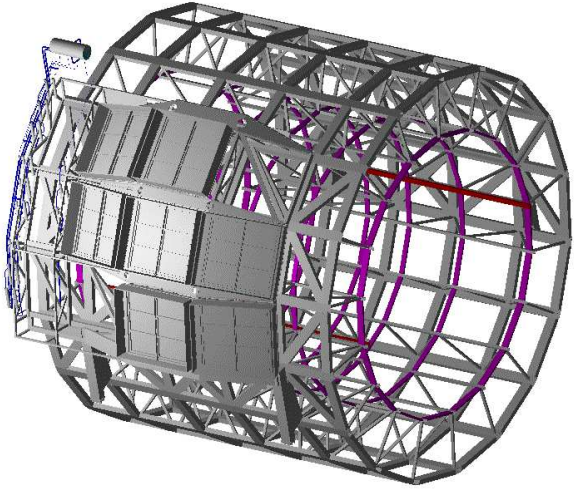}
        \caption{Layout of the HMPID modules on the support cradle as mounted on the ALICE space frame.}
        \label{fig:HmpidOnFrame}
\end{figure}

\section{Description of the installed HMPID}
\label{Sec:Det}
\subsection{The HMPID layout}
The High Momentum Particle Identification Detector consists
of seven identical proximity focusing type Ring Imaging Cherenkov Counter (RICH) modules.
The modules are located on an independent support cradle and mounted at the two o'clock
position of the ALICE space frame as shown in Fig.~\ref{fig:HmpidOnFrame}.
The seven modules (1.5 m $\times$ 1.5 m each) cover $-0.6$~$\leq \eta \leq$~0.6 in
pseudo-rapidity and 57.6$^{o}$ in azimuth. The chambers are tilted and positioned
in a cupola-like structure to focus on the nominal interaction point, at $\sim$~4.7~m distance from the modules.
The layout of the chambers on the support cradle maximizes the HMPID acceptance
for two-particle correlation studies at high-$p_{\rm T}$.

\subsection{Radiators and the liquid circulation system}

Fast charged particles, emerging from the collisions, create Cherenkov photons traversing the
low chromacity liquid $\rm C_{6}F_{14}$ radiators (with index of refraction of $n=1.2989$
at $\lambda=175$ nm).
The choice of the radiator liquid determines the momentum threshold; for liquid $\rm C_{6}F_{14}$
the momentum threshold is given by $p_{\rm th} = $ 1.21~$mc$, where $m$ is the particle mass.

Each HMPID module is equipped with three radiator vessels made of
NEOCERAM$^{\textregistered}$, providing 15~mm radiator thickness.
The NEOCERAM (a glass-ceramic material) is thermally compatible to
the fused silica plates, used as UV-transparent windows.

The radiator vessels are filled and emptied at a constant flow by the liquid circulation
system. To ensure safe remote operation the system is built on the gravity flow principle.
To minimize the contamination of water and oxygen (less than 5 ppm), and to achieve
the best transparency of the liquid $\rm C_{6}F_{14}$ radiator for Cherenkov photons,
filters are implemented for purification. The liquid circulation system is scheduled
to be completed and installed in its final location
at the beginning of 2008.
\subsection{The photo-detector}
The Cherenkov photons created in the liquid radiator refract out from the fused
silica windows and enter the proximity gap. The proximity volume is filled
with pure $\rm CH_{4}$ at ambient pressure and temperature.

The Cherenkov photons are detected on the pad cathodes of Multi-Wire Pad
Chambers (MWPC). Each module consists of 6 caesium iodide (CsI) covered
photocathodes, 60 cm $\times$ 40 cm each. Quantum efficiency $\sim$ 25\% (at
$\lambda=175$ nm) is achieved with the 300 nm thick CsI covered photocathodes.
The pad cathodes are segmented into 8 mm $\times$ 8.4 mm pads to provide
position sensitive information. Detailed description of the HMPID CsI
photocathode production can be found in~\cite{CsI}.

The total gas gain achievable is 5 $\times$ 10$^4$, with the cathodes kept
at ground and a positive voltage of 2050 V applied to the anode wires.

\begin{figure}[htbp]
    \centering
        \includegraphics[width=0.49\textwidth]{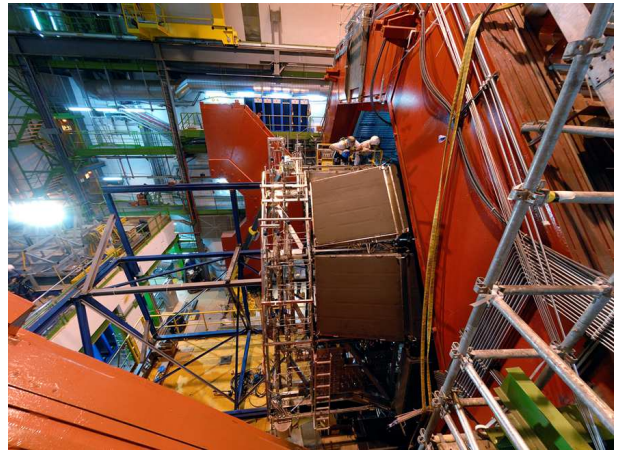}
        \caption{Installation of the HMPID modules in the ALICE magnet.}
        \label{fig:HmpidInstall}
\end{figure}
\subsection{Front End Electronics}
The front-end electronics is located at the back side of the HMPID modules.
The closure box is flushed with nitrogen and closed by a cooling panel, that removes the
dissipated heat of the electronics, 450 W per module.

The front-end electronics is based on two dedicated ASIC chips, GASSIPLEX
and DILOGIC~\cite{GassiplexDilogic}, processing the analogue signals from the pad cathodes.
The GASSIPLEX and the DILOGIC chips were successfully developed within the framework
of the HMPID project in the ALCATEL-MIETEC 0.7~$\mu$m technology.

The GASSIPLEX chip is a 16-channel analogue multiplexed low-noise signal processor.
Preceding the HMPID commissioning, the noise of the front-end chips and the map of dead or noisy pads
have been surveyed. The noise is found to be 1000\,e RMS with a dispersion of
less than 50\,e, and the number of dead/noisy pads is found to be 200 out of 161280 pads.

The DILOGIC chip is a sparse data scan readout processor used for zero suppression and pedestal
subtraction. Individual threshold and pedestal values can be processed for up to 64 channels.
Data from the 7 HMPID modules are readout via 14 standard ALICE optical links (DDL)~\cite{DDL}.

Digitization is performed using a commercial 12-bit ADC.  The digitization and
zero-suppression time does not depend on the occupancy, only on the multiplexing
rate of 3 chips (48 channels), which makes the event overlap negligible even in
the highest luminosity proton-proton collisions. Since particle identification
is not possible in the HMPID without momentum information, only events with TPC
information are of interest. The readout time after L2 trigger arrival (for the
estimated 12\% module occupancy) will be of the order of 300 $\mu$s, hence the
HMPID can perfectly cope with the readout rates foreseen for the TPC~\cite{DAQ}.

In total, 10080 GASSIPLEX and 3360 DILOGIC chips have been installed in the seven HMPID modules.

\section{Installation and commissioning}
\label{Sec:Inst}

All HMPID modules have been successfully installed on the supporting
cradle and moved into the ALICE magnet on 26 September 2006, as
shown in Fig.~\ref{fig:HmpidInstall}. Preceding the installation,
five HMPID modules were tested in test beam and two modules in
cosmic ray data taking~\cite{HmpidTestBeam}.
\begin{figure}[htbp]
    \centering
        \includegraphics[width=0.49\textwidth]{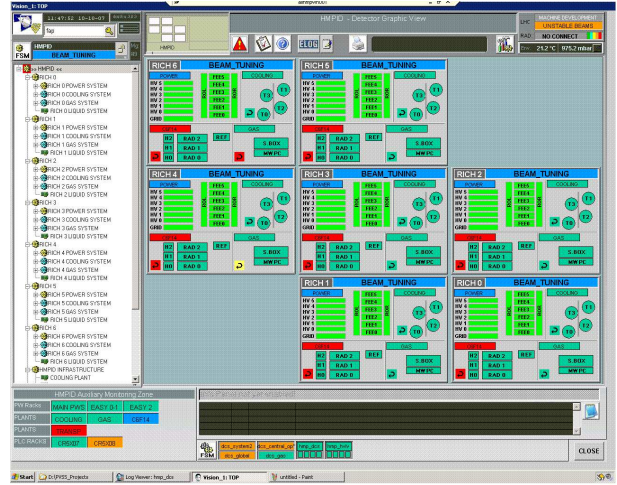}
        \caption{Detector-oriented view of the HMPID DCS control panel.}
        \label{fig:dcs1}
\end{figure}

While the HMPID is not operational, the modules are flushed with argon.
After the installation of the HMPID modules on the support cradle in the experimental
cavern and prior to moving the assembly into the
ALICE magnet, a leak was detected in one of the radiator trays (in module 6).
The loss of a single radiator (1 out of 21) represents $\sim$ 5\% loss of acceptance.
Due to the small impact on the physics potential of the detector
and the not negligible risks, including delays, involved in the repair,
it has been decided to install the detector inside the ALICE solenoid with 20 operational
radiators while flushing the leaking one with methane.
\begin{figure}[htbp]
    \centering
        \includegraphics[width=0.49\textwidth]{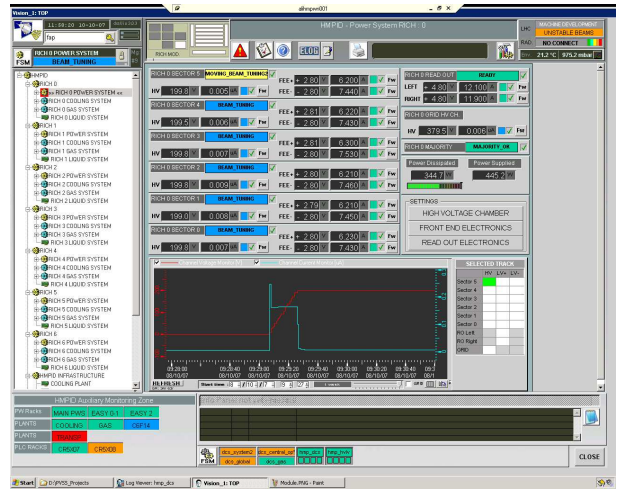}
        \caption{Device-oriented view of the HMPID Power Modules at High Voltage ramp up.}
        \label{fig:dcs2}
\end{figure}
%
%
%

The safe operation of the detector under commissioning and in normal operation is achieved
via the Detector Control System (DCS), developed within the HMPID project.
Since July 2007 all HMPID modules are fully powered and
controlled by DCS. Figure~\ref{fig:dcs1} shows the detector-oriented view of the fully powered
HMPID, with most subsystems operational. The HMPID DCS controls the subsystems of gas, cooling,
liquid circulation and transparency, high-voltage, low-voltage and the front end electronics as
shown in Fig.~\ref{fig:dcs1}. During the commissioning phase the alarm and interlock
levels have been set for the full scale cosmic test of ALICE. For expert operation, the DCS
provides access down to a single hardware element of the HMPID, in the device oriented view.
Figure~\ref{fig:dcs2} shows the device oriented view of the high voltage modules during
the high voltage ramp up of HMPID module 0.

\begin{figure}[thbp]
    \centering
        \includegraphics[width=0.49\textwidth]{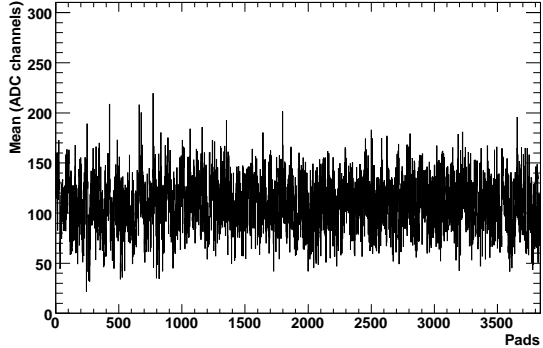}
     \caption{Average pedestal values of HMPID module 0 and photocathode 0, measured after the installation.}
        \label{fig:pedmean}
\end{figure}
\begin{figure}[htbp]
    \centering
        \includegraphics[width=0.49\textwidth]{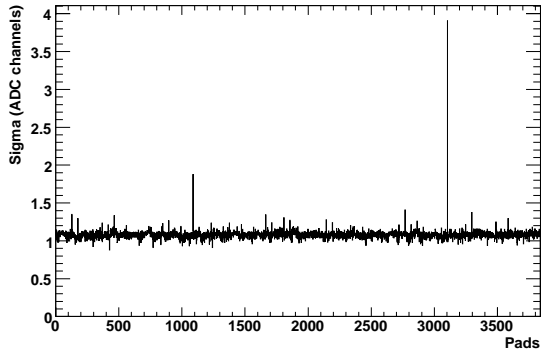}
     \caption{Average pedestal width of HMPID module 0 and photocathode 0, measured after the installation.}
        \label{fig:pedsigma}
\end{figure}

A long term stability test of the fully powered HMPID has been performed under
DCS control, archiving all relevant detector information in the central ALICE
database.  The continuous monitoring revealed a loose high voltage channel (in
module 5). The standard high voltage (2050 V) cannot be applied on this channel;
a lower high voltage setting is used.  This may only allow the detection of charged
particles, and not Cherenkov photons, in that high voltage sector.
This represents a $\sim$ 2.4\% (1 out of 42 high voltage channels) acceptance
loss.

After the long term stability test, the systematic survey of the read-out
electronics for noisy and dead channels has been repeated.  The first results of
the HMPID module 0 and photocathode 0 are shown in Figs.~\ref{fig:pedmean} and
\ref{fig:pedsigma}.  Figure~\ref{fig:pedmean} shows the mean pedestal value for
each photocathode pad.  The average pedestal value is found to be $\sim$ 110 out
of 4092 ADC channels.  Figure~\ref{fig:pedsigma} shows the width of the pedestal
distribution for each photocathode pad.  The average width is $\sim$ 1.1,
equivalent to a charge 1000\,e RMS, with small number of outlying pads.
This agrees with the expected value of the noise from the read-out electronics as
reported in~\cite{GassiplexDilogic}.

\section{Summary}

The as-built High Momentum Particle Identification Detector has been presented, along with
its commissioning and installation details.
A leak has been found in one of the 21 radiators and one loose high
voltage channel out of 42 high voltage channels has been discovered.
However, the physics goals of HMPID are not significantly
compromised by the small acceptance loss.
All seven HMPID modules were mounted on the support cradle and installed
in the ALICE magnet in September 2006. Full control of the HMPID modules
has been achieved via the Detector Control System in July 2007.
The liquid circulation system is scheduled to be completed at the beginning of
2008.
Hence, the HMPID will be ready for the data taking by the expected start of
the LHC in the summer of 2008.



\end{document}